\documentclass[aps,prl,twocolumn,english,showpacs,10pt]{revtex4-1} 
\usepackage{amsmath,bm}
\usepackage{amssymb}
\usepackage{graphicx}
\usepackage{babel}
\usepackage{cases}
\usepackage{natbib}
\usepackage{floatrow}
\usepackage{dblfloatfix}
\usepackage{xcolor}

\makeatletter

\newcommand{\Rmnum}[1]{\expandafter\@slowromancap\romannumeral #1@}
\makeatother

\begin{document}

\title{Subradiant Dimer Excitations of Emitter Chains Coupled to a 1D Waveguide}
\author{Yu-Xiang Zhang}
\email{iyxz@phys.au.dk}
\author{Chuan Yu}
\email{chuan@pks.mpg.de}
\altaffiliation[Present address: ]{Max Planck Institute for the Physics of Complex Systems, N\"{o}thnitzer Str. 38, 01187 Dresden, Germany}
\author{Klaus M{\o}lmer}
\email{moelmer@phys.au.dk}
\affiliation{Department of Physics and Astronomy, Aarhus University, 8000 Aarhus C, Denmark}
\date{\today}

\begin{abstract}
This Letter shows that chains of optical or microwave emitters coupled to 
a 1D waveguide support subradiant states with close pairs of excited emitters, 
which have longer lifetimes than even the most subradiant states with only a 
single excitation. Exact, analytical expressions for non-radiative excitation 
dimer states are obtained in the limit of infinite chains. To understand the 
mechanism underlying these states, we present a formal equivalence between 
subradiant dimers and single localized excitations around a chain defect (unoccupied site). 
Our analytical mapping permits extension to emitter chains coupled to the 3D free space vacuum field.
\end{abstract}
\maketitle

Subradiance, the cooperative inhibition of spontaneous emission from an ensemble of emitters,
has been pursued since the seminal work by Dicke \cite{Dicke1954} and has been observed only recently
in atomic gases \cite{Guerin2016,Weiss2018} and in metamaterial arrays \cite{Jenkins2017}.
Applications in quantum information processing \cite{Paulisch2016}
motivate the studies of collective light-matter interactions, including the subradiant excitations of 
one-dimensional (1D) emitter chains \cite{Haakh2016,Ruostekoski2016,Ruostekoski:2017aa,Zoubi:2014aa,
Kornovan2016,Sutherland2016,Olmos:2013aa,Bettles:2016aa,Jen:2016aa,Asenjo-Garcia2017,Albrecht2018,
Henriet:2019aa,Zhang:2019aa,Solano:2017aa,Paulisch2016,Asenjo-Garcia:2019aa,Kornovan2019aa,Zhang:2019ab,Wang:2018aa},
2D arrays \cite{Perczel2017,Facchinetti:2016aa,Guimond:2019aa} and other geometries \cite{Needham2019aa,Moreno-Cardoner:2019aa,Lee:1973aa}.
The phenomenon of subradiance is found to occur due to different mechanisms,
e.g., spin waves with wave numbers outside the ``light line'' \cite{Asenjo-Garcia2017},
entangled states between remote ensembles \cite{Guimond:2019aa,Needham2019aa,Moreno-Cardoner:2019aa},
subradiant edge states enabled by nontrivial topology \cite{Perczel2017,Zhang:2019ab,Wang:2018aa}, etc. However, these results were so far restricted to ensembles with only a single excitation while subradiant states with more excitations have remained largely unexplored.

An exception is the so-called fermionic multi-excitation
subradiant states in 1D systems \cite{Albrecht2018,Asenjo-Garcia2017,Henriet:2019aa}.
While one might expect emitter saturation to play only a perturbative role in the few-excitation scenarios \cite{Porras:2008aa}, 
it enforces an equivalence between the multi-excitation subradiant states 
and the Tonks-Girardeau gas of hard core bosons \cite{Zhang:2019aa}.
This suggests a class of subradiant states with state amplitudes which are anti-symmetric 
combinations of the one-excitation subradiant states.

However, numerical analyses of emitter chains coupled to a 1D waveguide 
reveal the existence of another family of subradiant states with entirely different properties, see \cite{Zhang:2019aa} and \cite[Sec.~A]{sp}.
In this Letter, we first numerically demonstrate and assess the extraordinary properties of
subradiant states with very close pairs of excited emitters, i.e., subradiant dimers. In particular, we find that for
specific distances between the excited emitters their radiative lifetimes can be longer than
the fermionic states and even than the most long-lived one-excitation states \cite{:aa}.
Then, we present an analytical treatment that explains the \emph{confinement} mechanism that leads to the
subradiant dimers by a mapping to the localized subradiant excitation near an unoccupied site (defect)
in the chain. This confinement-localization mapping is valid under more general conditions and allows extension of our
analysis, e.g., to emitter chains coupled to the 3D free space quantized field.

\begin{figure}[b]
  \centering
    \includegraphics[width=0.97\textwidth]{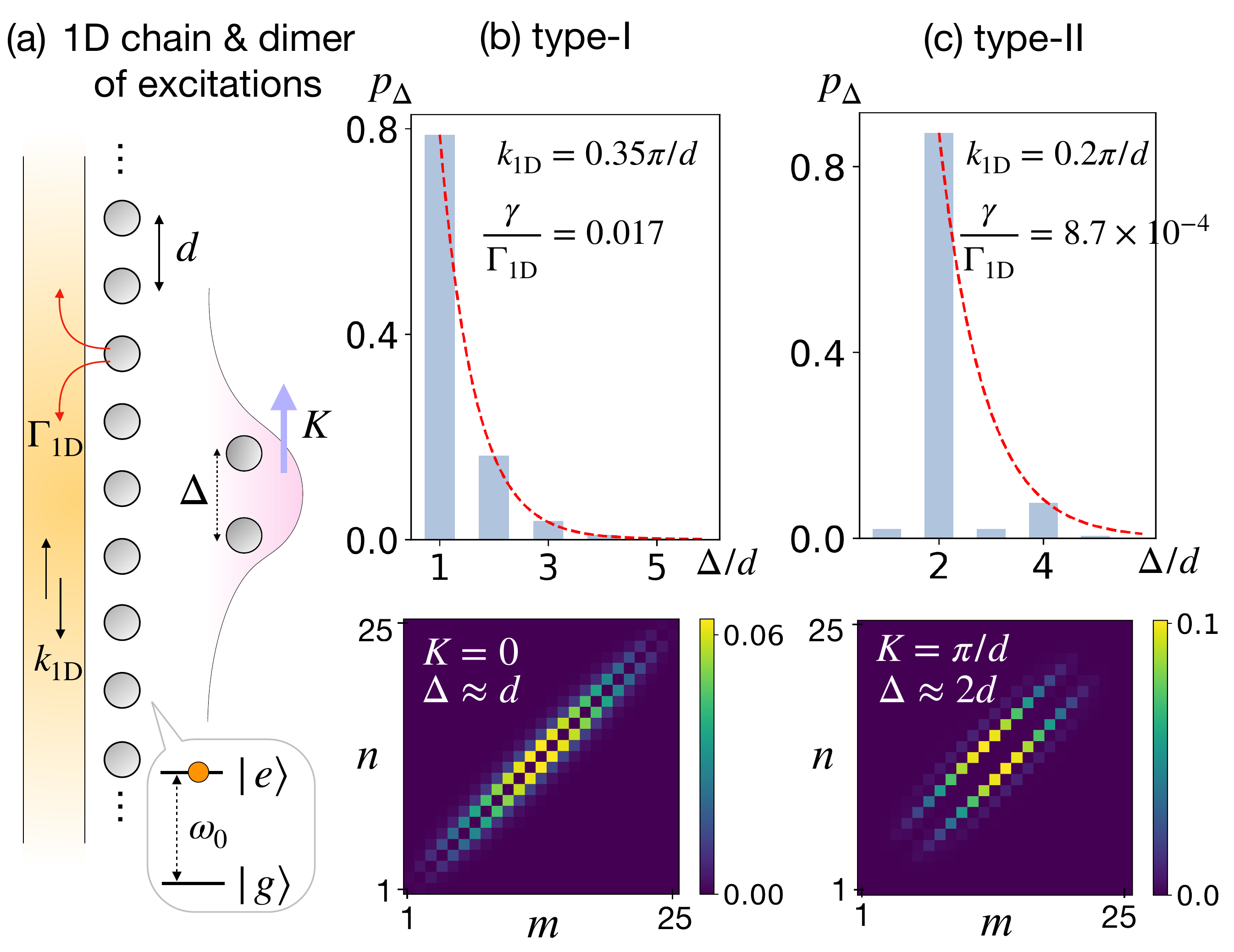}
\caption{(a) A chain of two-level emitters couple to a 1D waveguide with parameters introduced in the main text.
In the limit of infinite chains, a dimer state has two excitations characterized by a short relative distance $\Delta$
and the delocalized ``center of mass'' (corresponding to a well defined total wave number $K$).
In this Letter we study two types of subradiant dimers: type-\Rmnum{1} states, dominated by $K=0$ and $\Delta=d$,
and type-\Rmnum{2} states dominated by $K=\pi/d$ and $\Delta=2d$.
For finite chains of N=25 emitters, (b,c) show examples of type-\Rmnum{1} and type-\Rmnum{2} dimers.
Parameters of the resonant wave number and single emitter decay rates are specified in the upper
panels which show the distribution of the separation $\Delta$ between the excitations. The lower panels show
the strongly correlated spatial distribution $|\langle m, n|\psi\rangle|^2$ of excited emitter pairs in the dimer states. }
\label{fig1}
\end{figure}

\paragraph*{Spin Model.} Consider a chain of $N$ two-level emitters equally spaced by the distance $d$, as illustrated in Fig.~\ref{fig1}(a).
Each emitter has a ground state $|g\rangle$ and an excited state $|e\rangle$ with transition frequency $\omega_0$. The
emitters are coupled to a 1D waveguide that supports light modes with a linear dispersion relation.
Using the Born-Markov approximation, the waveguide modes can be
eliminated to yield an effective theory for the emitters \cite{Dung2002},
which entails a non-Hermitian infinite-range spin-spin interaction Hamiltonian \cite{Chang2012,Caneva:2015aa,Roy:2017aa}
\begin{equation}\label{Heff}
H_{\mathrm{eff}}=-\frac{i}{2}\Gamma_{\text{1D}}\sum_{m,n=1}^{N}e^{ik_{\text{1D}}|z_m-z_n|}\;\sigma_{m}^{\dagger}\sigma_{n}.
\end{equation}
The bare excitation energies of the individual emitters are not included,
$\Gamma_{\text{1D}}$ is the decay rate of an individual emitter coupled to the waveguide, $k_{\text{1D}}$ is the
wave number of the waveguide mode resonant with $\omega_0$, $z_m$ is the position coordinate of the $m^{th}$ emitter and
$\sigma_m^{\dagger}=|e\rangle_m\langle g|$.
In a perfect experimental implementation of \eqref{Heff} $\Gamma_{\text{1D}}$ should dominate all other decay processes, as in photonic crystal waveguides \cite{Zang:2016aa,Chang:2018aa} and superconducting qubits coupled to
transmission lines \cite{Astafiev:2010aa,Hoi:2011aa,VanLoo2013,Mirhosseini:2019aa}.

In the Monte Carlo wave function formalism \cite{Moelmer1993}, the state of the emitter chain evolves under Eq.~\eqref{Heff}, interrupted by stochastic quantum jumps
representing spontaneous emission of a photon. The jump rate makes a system prepared
in a \emph{right eigenstate} of $H_{\text{eff}}$ maintain its excitation with a
probability that decays with twice the negative imaginary part of the corresponding eigenvalue.
In this work,  we obtain these eigenstates by the exact diagonalization of $H_{\text{eff}}$ with use of the
SLEPc (Scalable Library for Eigenvalue Problem Computations) \cite{slepc2005}.

\paragraph*{Subradiant dimer excited states.}
We focus on the two-excitation subspace of eigenstates of $H_{\mathrm{eff}}$, for lattices with $0<k_{\text{1D}}d<\pi/2$.
This Hilbert space is spanned by states $|n, m\rangle$ where the $m^{th}$ and $n^{th}$ emitters are excited.
As we illustrate in Fig.~\ref{fig1}, by numerical diagonalization of $H_{\text{eff}}$ we find subradiant dimer states with delocalized center of mass $Z_c=(z_m+z_n)/2$ and well-defined distance $\Delta=|z_m-z_n|$ between the excitations.
To understand the appearance of these states and their properties, we introduce basis states,
\begin{equation}\label{K-Delta-state}
|K;\Delta\rangle=\sum_{Z_c=z_1+\Delta/2}^{z_N-\Delta/2}e^{iK Z_c}|Z_c-\frac{\Delta}{2}, Z_c+\frac{\Delta}{2}\rangle,
\end{equation}
with {\it center-of-excitation} wave number $K$ and spatial separation $\Delta$ between the excitations.
In a finite chain, $K$ is not conserved but $Z_c$ distributions resembling standing waves appear due to the boundary conditions at the chain ends.
The expansion of the identified type-\Rmnum{1} dimers on the basis states \eqref{K-Delta-state} have wave numbers $K\approx 0$.
For infinite chains, $K$ becomes a good quantum number and the type-\Rmnum{1} 
dimers can expressed as $ \sum_{\Delta}e^{iq_{\text{\Rmnum{1}}}\Delta/d}|K,\Delta\rangle$,
where $q_{\text{\Rmnum{1}}}d=-i\ln\cos(k_{\text{1D}}d)$, see \cite[Sec.~B]{sp}. This implies a probability
distribution for the separation $\Delta > 0$ between the excitations
\begin{equation}\label{dimer1}
p_{\text{\Rmnum{1}}}(\Delta)\propto (\cos k_{\text{1D}} d)^{2\Delta/d}
\end{equation}
with the dominant amplitude on $\Delta=d$, see Fig.~\ref{fig1}(b).
For the type-\Rmnum{2} dimers identified, $K\approx \pi/d$,
and on an infinite chain the amplitudes on odd values of $\Delta/d$ vanish and the state can be expressed as
$\sum'_{\Delta}e^{iq_{\text{\Rmnum{2}}}\Delta/d}|K,\Delta\rangle$.
The summation includes only the even values of $\Delta/d$, and $q_{\text{\Rmnum{2}}}d=[\pi+i\ln\cos(2k_{\text{1D}}d)]/2$.
This implies a distribution for even valued separations
\begin{equation}\label{dimer-2}
p_{\text{\Rmnum{2}}}(\Delta)\propto (\cos 2k_{\text{1D}} d)^{\Delta/d}
\end{equation}
with the dominant amplitude on $\Delta=2d$, see Fig.~\ref{fig1}(c). These dimers are perfectly subradiant with vanishing decay rates on infinite chains, i.e., the corresponding eigenvalues of $H_{\mathrm{eff}}$ are real.
In \cite[Sec.~B]{sp}, we derive the asymptotic eigenvalues
of the two types of dimers, viz., $\omega_{\text{\Rmnum{1}}}=2\Gamma_{\text{1D}}\cot(k_{\text{1D}}d)$ and
$\omega_{\text{\Rmnum{2}}}=2\Gamma_{\text{1D}}\cot(2k_{\text{1D}}d)$, and numerically obtain their corrections on a finite chain.
Knowing these asymptotic values allows efficient search for the eigenstates for a finite chain with a large number of
emitters by the Krylov-Schur algorithm \cite{stewart2002} with the shift-and-invert method \cite{saad2011}.

\begin{figure}[t]
  \centering
    \includegraphics[width=\textwidth]{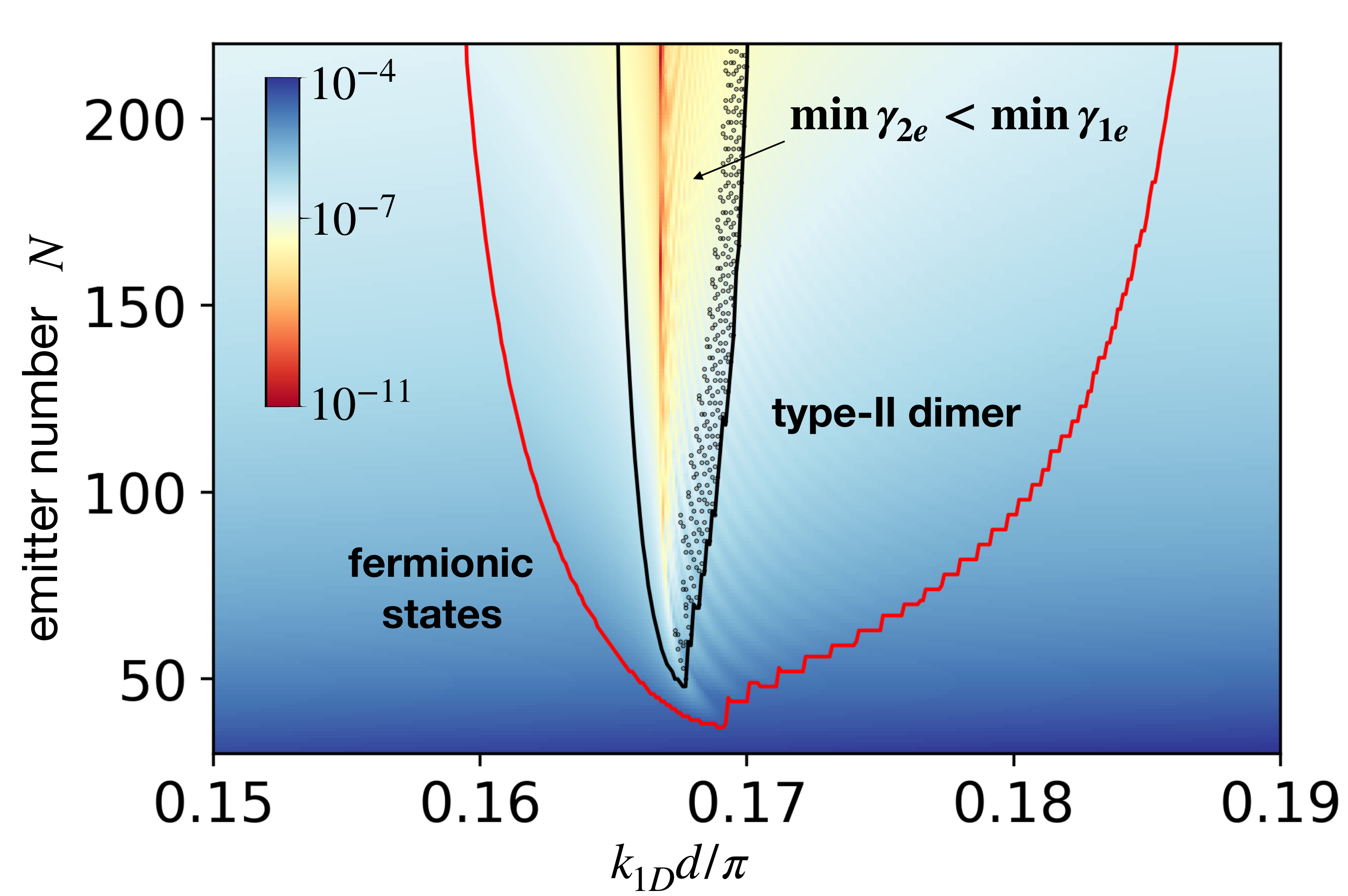}
\caption{Decay rates of the most subradiant type-\Rmnum{2} dimers (shown by colors, with unit $\Gamma_{\text{1D}}$) as function of emitter number $N$ and lattice distance $d$.
The red curve encloses the regime  where the type-\Rmnum{2} dimers are longer lived than the fermionic states.
The black line encloses the regime where the type-\Rmnum{2} dimers are even longer lived than the one-excitation states. In the shaded region, the dimer decay rate oscillates around the single excitation rates as function of $N$and $d$.}
\label{phase}
\end{figure}

\begin{figure}[b]
  \centering
    \includegraphics[width=\textwidth]{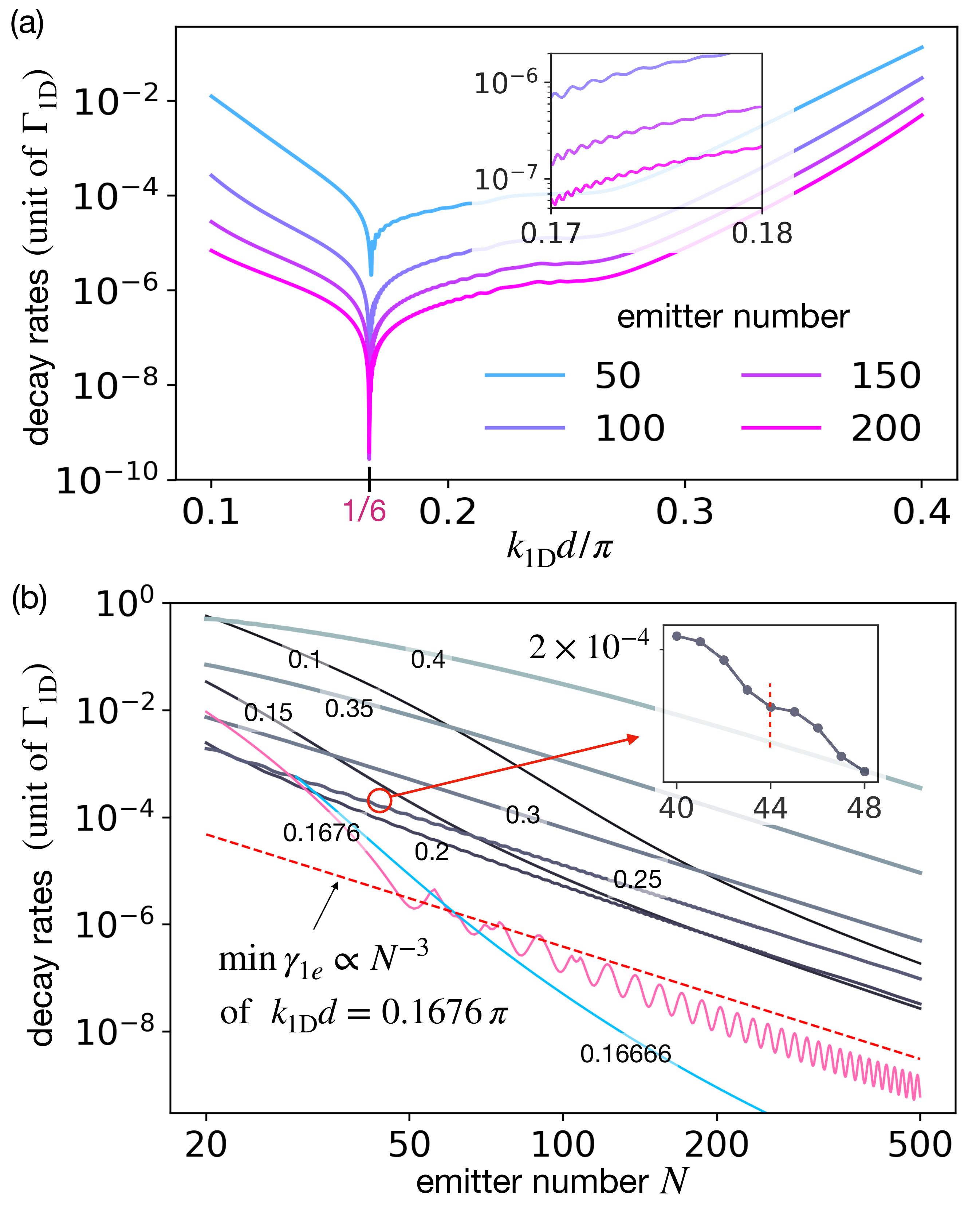}
\caption{Decay rates of the most subradiant type-\Rmnum{2} dimer states.
(a) Decay rates as function of $k_{\text{1D}}d/\pi$ for different values of $N$. A narrow dip is found near $k_{\text{1D}}d=\pi/6$. The
insert shows a magnified view of the oscillatory behavior of the decay rates.
(b) Decay rates as function of emitter number for $k_{\text{1D}}d/\pi$ from 0.1 to 0.4 (every 0.05), 0.1676 and 0.16666.
The dashed line shows the decay rate of the most subradiant one-excitation state for $k_{\text{1D}}d=0.1676\pi$, scaling as $N^{-3}$.
The insert shows a period four  modulation of the decay rate as function of $N$ for $k_{\text{1D}}d=0.25\pi$.}
\label{decay-N-k}
\end{figure}

The dimers on finite chains have small but finite decay rates.
We find that the minimal decay rates of the most subradiant type-\Rmnum{1} states
scale asymptotically as nearly $N^{-2}$, see \cite[Sec.~C]{sp}. However, the type-\Rmnum{2} states show longer lifetimes and a more complex behaviour.
The decay rate of the most subradiant type-\Rmnum{2} dimers is shown versus
$N$ and $k_{\text{1D}}d$ in Fig.~\ref{phase}, where a narrow region between $k_{\text{1D}}d=0.16\pi$ and $0.17\pi$
is distinctively subradiant. Moreover, fringe textures are seen to the right hand side of that region.
We compare the minimal decay rates of type-\Rmnum{2} dimers with those of the most subradiant
fermionic states and one-excitation states, and obtain the critical parameters of $k_{\text{1D}}d$ and $N$ shown by
the red and black boundary curves in Fig.~\ref{phase}. Specifically, Fig.~\ref{phase} demonstrates
that the type-\Rmnum{2} dimer can be more subradiant than the fermionic states, and even the one-excitation states.
This observation disproves an unwritten orthodoxy that states with more excitations have shorter lifetimes,
being true, for example, for the fermionic states that decay with the sum
of their one-excitation constituent decay rates \cite{Asenjo-Garcia2017,Albrecht2018,Henriet:2019aa}.
Fig.~\ref{phase} shows that a short chain with $N=48$ emitters is sufficient to observe the even more subradiant dimer states,
in the case of $k_\text{1D}d = 0.1676\pi$.

To study the extremely subradiant region in more detail, we plot decay rates of the most subradiant type-\Rmnum{2} dimers
for a few values of $N$ in Fig.~\ref{decay-N-k}(a). A sharp dip in decay rates appears around $k_{\text{1D}}d=\pi/6$ and a magnified view of the
interval $k_{\text{1D}}d/\pi\in[0.17,0.18]$ shows the fringe textures observed in Fig.~\ref{phase}. 
The robustness of the results to position disorder is discussed in \cite[Sec.~D]{sp}.
In Fig.~\ref{decay-N-k}(b),  we observe different scalings of the decay rate with $N\leq 500$ for different 
values of $k_{\text{1D}}d$. Around $k_{\text{1D}}d=\pi/6$, the decay rates thus fall off faster than $N^{-3}$ and 
show oscillations breaking the conventional monotonicity with $N$. For $k_{\text{1D}}d$ near $0.25\pi$, 
the decay rate is weakly modulated with a period of 4, see insert in Fig.~\ref{decay-N-k}(b): adding 4 
emitters makes the chain longer by half a resonant wavelength.

When $k_{\text{1D}}d=0.25\pi$, Eq.~(\ref{dimer-2}) vanishes so that amplitudes of $\Delta>2d$ are completely
suppressed. On infinite chains this state is an eigenstate of both the centre-of-excitation wave number, 
equivalent to a total momentum, $\hat{p}_{1}+\hat{p}_{2}$, and the relative position coordinate,
$\hat{x}_{1}-\hat{x}_{2}$, i.e., it is an implementation of the Einstein-Podolsky-Rosen state \cite{Einstein:1935aa}.
A similar state is found for the type-\Rmnum{1} dimer for $k_{\text{1D}}d=0.5\pi$.

\paragraph*{Confinement-localization mapping.}
The Hamiltonian Eq.~\eqref{Heff} does not provide any direct evidence for the emergence of the subradiant dimer states, and neither does our analysis based on the Holstein-Primakoff transformation \cite{Zhang:2019aa}.
Here we provide the physical mechanism leading to the long lived excitation dimers.

Applying $H_{\text{eff}}$ on the ansatz of Eq.~(\ref{K-Delta-state}) yields
\begin{equation}\label{H-K-delta}
H_{\mathrm{eff}}|K;\Delta\rangle=\sum_{\Delta'\geq d}\mathcal{H}^K_{\Delta,\Delta'}|K;\Delta'\rangle+(\text{tails}).
\end{equation}
It separates $H_{\text{eff}}$ into contributions preserving $K$, i.e.,
the matrix $\mathcal{H}^{K}$ defined by elements
\begin{equation}\label{H_K}
\mathcal{H}^{K}_{\Delta,\Delta'}=-\frac{i}{2}\Gamma_{\text{1D}}\sum_{\epsilon,\epsilon'=\pm1}
e^{i(k_{\text{1D}}+\epsilon\frac{K}{2})|\Delta-\epsilon'\Delta'|},
\end{equation}
and remaining terms, denoted by ``tails'' that break the conservation of $K$, see \cite[Sec.~E]{sp}.
The ``tails'' vanish when $N\rightarrow\infty$. Thus the Hamiltonian
$\mathcal{H}^{K}$ acting on the relative position eigenstates is essential for the formation of dimers and must explain their vanishing decay rates
on infinite chains.

Our key insight is that
the eigenstates of $\mathcal{H}^{K}$ can be uniquely mapped to the even-parity eigenstates of a
Hamiltonian $\mathcal{H}^K_{\text{def}}$ with matrix elements
\begin{equation}\label{defect model}
(\mathcal{H}^{K}_{\text{def}})_{\Delta,\Delta'}=-\frac{i}{4}\Gamma_{\text{1D}}
\sum_{\epsilon=\pm 1}e^{i(k_{\text{1D}}+\epsilon\frac{K}{2})|\Delta-\Delta'|},
\end{equation}
where $\Delta/d,\Delta'/d$ attain both positive and negative values $\{\pm1, \pm2,\cdots \}$.
As shown in \cite[Sec.~F]{sp},
for an eigenstate $|\psi\rangle$ of $\mathcal{H}^{K}$ (with eigenvalue $\lambda$),
there is a corresponding even-parity eigenstate $|\psi_{\text{def}}\rangle$ (unnormalized)
of $\mathcal{H}^{K}_{\text{def}}$ with the eigenvalue $\lambda/2$ that satisfies
$\langle\Delta|\psi_{\text{def}}\rangle=\langle\Delta|\psi\rangle$ for indices $\Delta>0$.

\begin{figure}[b]
  \centering
    \includegraphics[width=\textwidth]{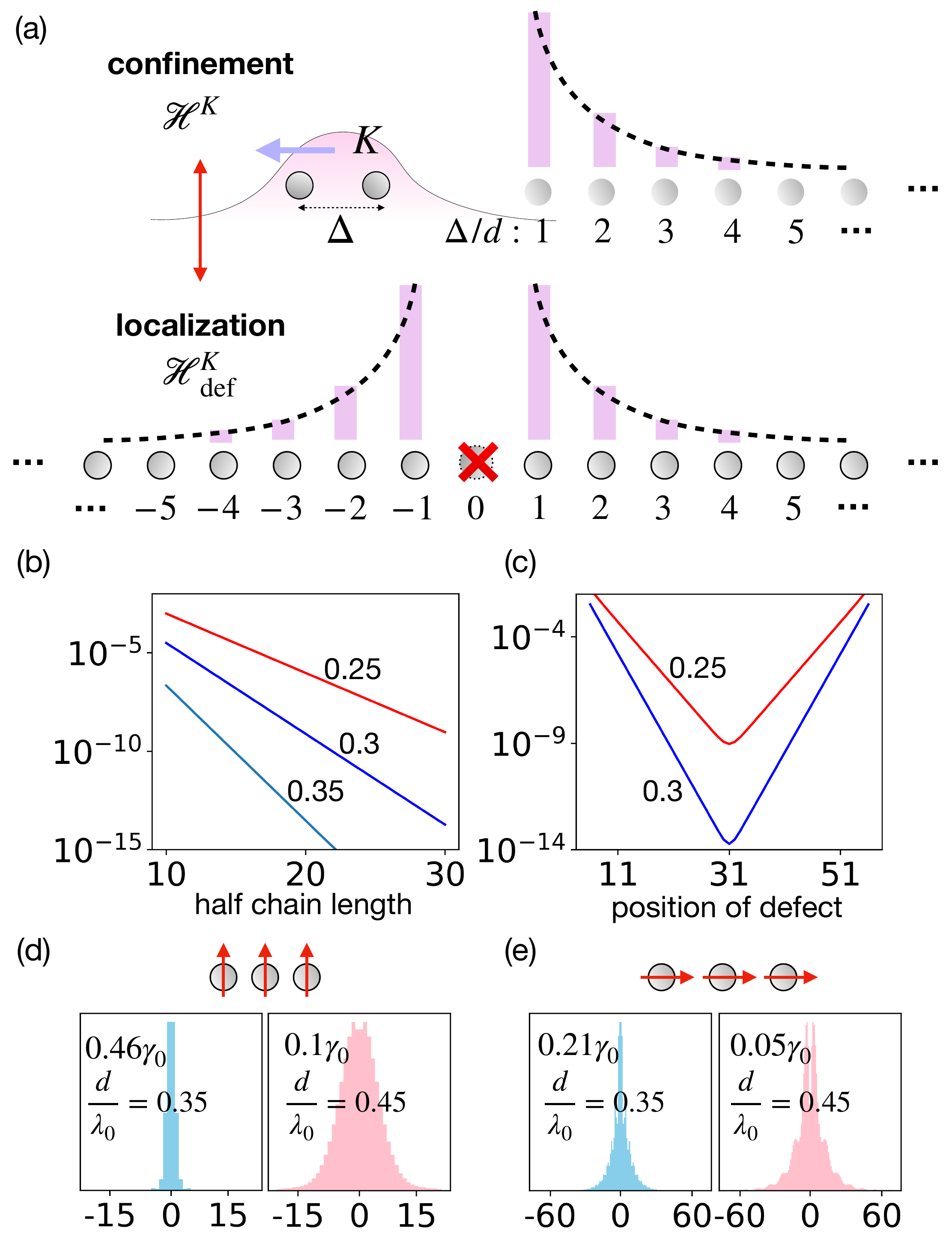}
\caption{(a) The confinement of excitation pairs is equivalent to localization of
an excitation around a missing site. (b) Decay rates (in units of $\Gamma_{\text{1D}}$)
of the state localized around the central missing site, for $k_{\text{1D}}d/\pi=0.25, 0.3, 0.35$.
(c) Decay rates of the localized state as function of the position of the missing site in a chain with 60 emitters
and $k_{\text{1D}}d/\pi=0.25, 0.3$. Profiles of the localized states of emitter chains coupled to free quantized field in 3D, with
(d) transverse and (e) parallel polarization, for $d=0.35$ and $0.45$ times
the resonant wavelength $\lambda_0$. The decay rates are given in units
of the single emitter spontaneous emission rate $\gamma_0$. }
\label{fig4}
\end{figure}

The above mapping implies that, the dimer state is
equivalent to an eigenstate of the Hamiltonian $\mathcal{H}^{K}_{\text{def}}$, describing a single 
excitation localized around a defect (unoccupied site) at $\Delta=0$, as illustrated in Fig.~\ref{fig4}(a).
Actually for the case of $K=0$, $\mathcal{H}_{\text{def}}^{K=0}$ is just the defect version of
Eq.~\eqref{Heff}. The localized defect modes have analytical solutions elaborated in \cite[Sec.~G]{sp}.
On an infinite chain, the localized eigenstate of $\mathcal{H}_{\text{def}}^{K=0}$ can be written
as $\propto \sum_{\Delta}e^{iq_{\text{\Rmnum{1}}}|\Delta|}|\Delta\rangle$ with $q_{\text{\Rmnum{1}}}d=-i\ln\cos(k_{\text{1D}}d)$.
This is in agreement with the spatial factor Eq.~\eqref{dimer1} found for the type-\Rmnum{1} dimers.
The eigenvalue of $\mathcal{H}_{\text{def}}^{K=0}$ for the localized state is $\Gamma_{\text{1D}}\cot(k_{\text{1D}}d)$,
exactly half of that of the type-\Rmnum{1} dimer $\omega_{\text{\Rmnum{1}}}$, as predicted by
the confinement-localization mapping.
The case of $K=\pi/d$ (type-\Rmnum{2} dimers)
is equivalent to the case of $K=0$ \cite[Sec.~H]{sp}, up to alternating sign flips and the replacement of $d\rightarrow 2d$.
This equivalence explains the resemblance between Eqs. \eqref{dimer1} and \eqref{dimer-2}, and 
between the eigenvalues $\omega_{\text{\Rmnum{1}}}$ and $\omega_{\text{\Rmnum{2}}}$.

The excitation of an emitter blocks further excitation and the dimer state is stable because each excitation serves as a defect
supporting the localization of the other one. 
For finite chains, as shown in Fig.~\ref{fig4}(b) the localized state around a defect
has a decay rate suppressed \emph{exponentially} in the emitter number $N$, and if the defect is not at the chain center,
the decay rate is determined by the length of the shorter subchain, see Fig.~\ref{fig4}(c). This dependence is much faster than the $N^{-3}$ scaling on
chains free from defects \cite{Asenjo-Garcia2017,Albrecht2018,Henriet:2019aa,Zhang:2019aa} and
it comes about because the localized state is a superposition of a left and a right
excited subchain. Their destructive interference results in the extreme subradiance, seen also in
Refs. \cite{Guimond:2019aa,Needham2019aa,Moreno-Cardoner:2019aa}.

The finite size effects shown in Figs. \ref{phase} and \ref{decay-N-k} are due to the ``tails'' of Eq.~\eqref{H-K-delta}.
Since the states $|K,\Delta\rangle$ have finite width of $\Delta$, the ``tails''
is restricted to a short section of length $\Delta$ at each end, but their complicated expression \cite[Sec.~E]{sp} 
prevents analytical solution. We may, however, infer that the dimers become seriously affected at the chain ends, 
and hence an interplay between the length of the chain and the bond-length
of the dimer may be responsible for the periodic oscillations seen in Fig.~\ref{decay-N-k}(b) and the 
fringe texture seen in Fig.~\ref{phase}. The strong dependence of the decay rates on the emitter 
separation $d$ might be equivalent to the observation, see Ref.~\cite{Kornovan2019aa} of special emitter distances 
leading to extraordinary (single excitation) subradiant states.

\paragraph*{Universality.}
The mapping between confinement and localization can be extended by linearity to
Hamiltonians written as $H_{\text{eff}}=\int d\mu(k_{\text{1D}})H_{\text{eff}}(k_{\text{1D}})$, where $d\mu(k_{\text{1D}})$ is an integral
measure over the variable $k_{\text{1D}}$ and $H_{\text{eff}}(k_{\text{1D}})$ refers to Eq.~\eqref{Heff}.
This for example covers 1D emitter chains coupled with 3D free space modes where $\mu(k_{\text{1D}})$ is given in Ref.~\cite{Zhang:2019aa}.
Also here, localized subradiant states will exist, and we show their excitation amplitude in Fig.~\ref{fig4}(d,e) 
(see phase profiles in \cite[Sec.~I]{sp}),
for emitters polarized both transverse and parallel to the chain.
In contrast to the 1D waveguide, these localized states have finite decay rates in the infinite chain limit.
Due to the mapping, we can conclude that subradiant excitation dimers exist and that 
they have intrinsic finite decay rates also in the limit of infinite chains.

\paragraph*{Conclusion and discussion.}
In this Letter, we have introduced subradiant excited dimers
of emitter chains coupled to a 1D waveguide. We showed that such (type-\Rmnum{2}) dimers 
can be more subradiant than even the longest lived one-excitation states of the system. 
Their decay rates show unusual dependence, including non-monotonic
wiggles, as function of $N$ when $k_{\text{1D}}d$ is slightly larger than $\pi/6$.
We identify the intrinsic emitter saturation and the resulting emergence of mutual
defect modes as the cause of the long lived dimer states, and we
use a rigorous mapping between confinement and localization to obtain their physical properties. 
This mapping is valid under general conditions and gives access to the properties of similar 
states of emitter chains decaying into 3D free space field modes.

We propose to verify the predictions experimentally, and to address single excited states around defects, 
e.g., by exciting emitters around a missing or suitably perturbed site, and waiting for 
excitation amplitudes on orthogonal excitation modes to decay. 
To verify the dimer subradiant states, one would excite a system uniformly, 
but one may exploit interactions to facilitate correlated excitation of dimers within 
certain distance ranges \cite{Douglas:2016aa}, and thus maximize the overlap with the 
long lived states identified in this Letter. Other efficient ways to couple the bound 
states or localized states may be mediated by
ancillary emitters distributed off the 1D lattice sites \cite{Mirhosseini:2019aa,Gonzalez-Tudela:2015aa}.
Finally, we imagine that our method to formally map doubly excited states on localized 
states around defects may become a useful ingredient in other lattice models and contribute 
to the analysis of other quasiparticle confinement phenomena, cf., recent findings 
in Ising spin chains with long-range interactions \cite{Liu:2019aa}.

\begin{acknowledgments}
This work was supported by the Villum Foundation, the European Unions Horizon 2020
research and innovation program (Grant No. 712721, NanOQTech), and
the EuropeanUnion FETFLAG program (Grant No. 820391, SQUARE).
The numerical results were obtained at the Center for Scientific Computing, Aarhus University.
\end{acknowledgments}

\clearpage
\section{Supplemental Material}
\setcounter{equation}{0}
\setcounter{figure}{0}

\subsection{A. Distributions of the eigenvalues in the two-excitation sector}
\begin{figure}[h]
  \centering
    \includegraphics[width=\textwidth]{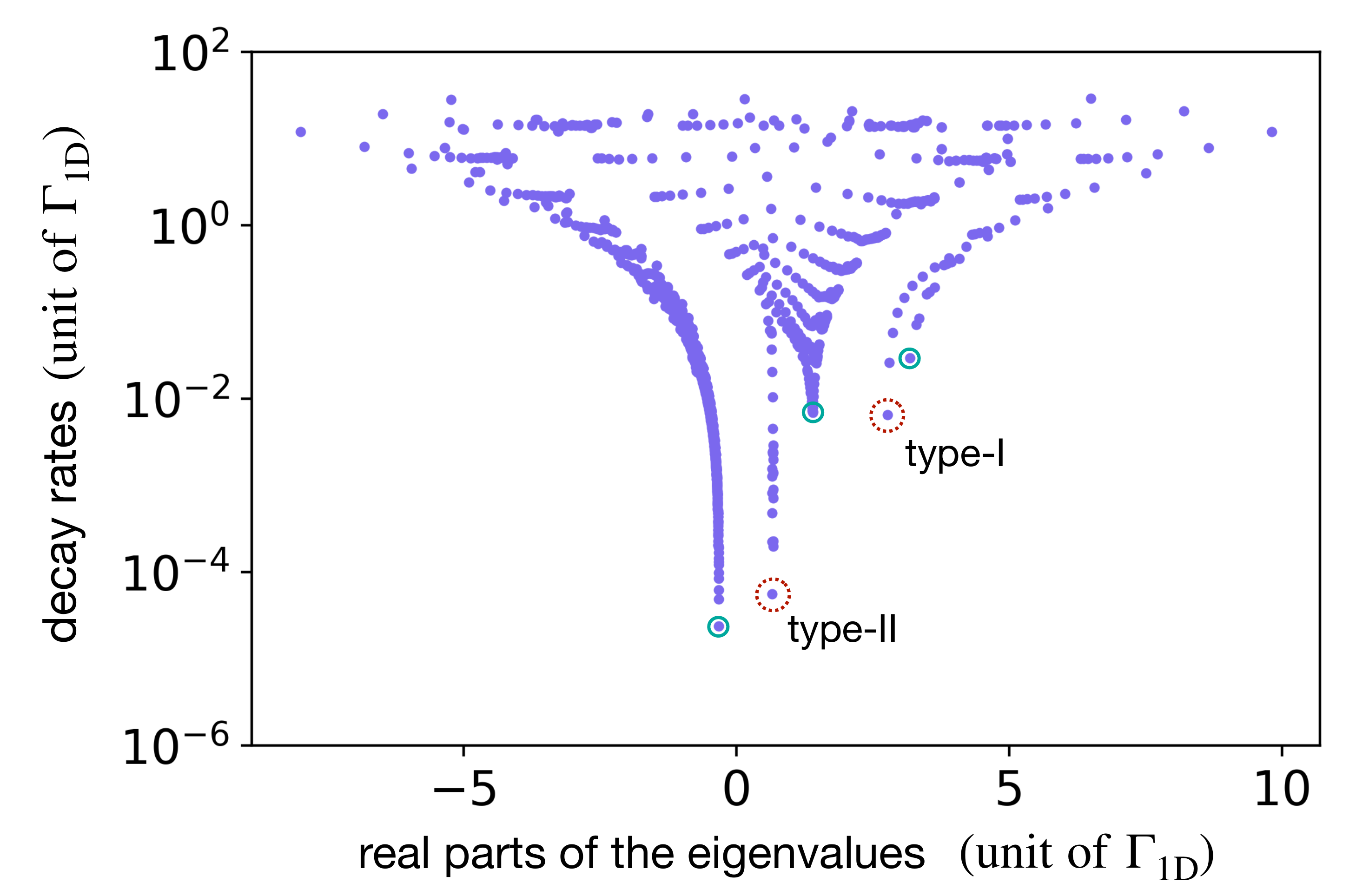}
\caption{The eigenvalues of the two-excitation sector $H_{\text{eff}}$
with $N=50$ and $k_{\text{1D}}d=0.2\pi$.
Five branches of subradiant states are identified.
Three of them are the fermionic states, of which the most subradiant ones are marked by green solid circles.
The other two are the dimers. The most subradiant type-\Rmnum{1} and type-\Rmnum{2} are marked by red dotted circles.}
\label{eigenvalues}
\end{figure}

The fermionic states and the dimers constitute all the subradiant states
in the two-excitation sector. This is verified by Fig.~\ref{eigenvalues}, where
we show all the eigenvalues of $H_{\text{eff}}$ for a chain of $N=50$ emitters coupled
to a 1D waveguide with  $k_{\text{1D}}d=0.2\pi$. In Fig.~\ref{eigenvalues} five branches of subradiant states
can be clearly identified.
For the three branches of fermionic states, their asymptotic eigenvalues can be obtained 
from the supplemental material (Sec.~A) of Ref. \cite{Zhang:2019aa}.
The asymptotic eigenvalues of the dimers are given in the main text and will be derived below in Sec.~B.

\subsection{B. Asymptotic eigenvalues of the dimers}
\begin{figure}[h]
  \centering
    \includegraphics[width=\textwidth]{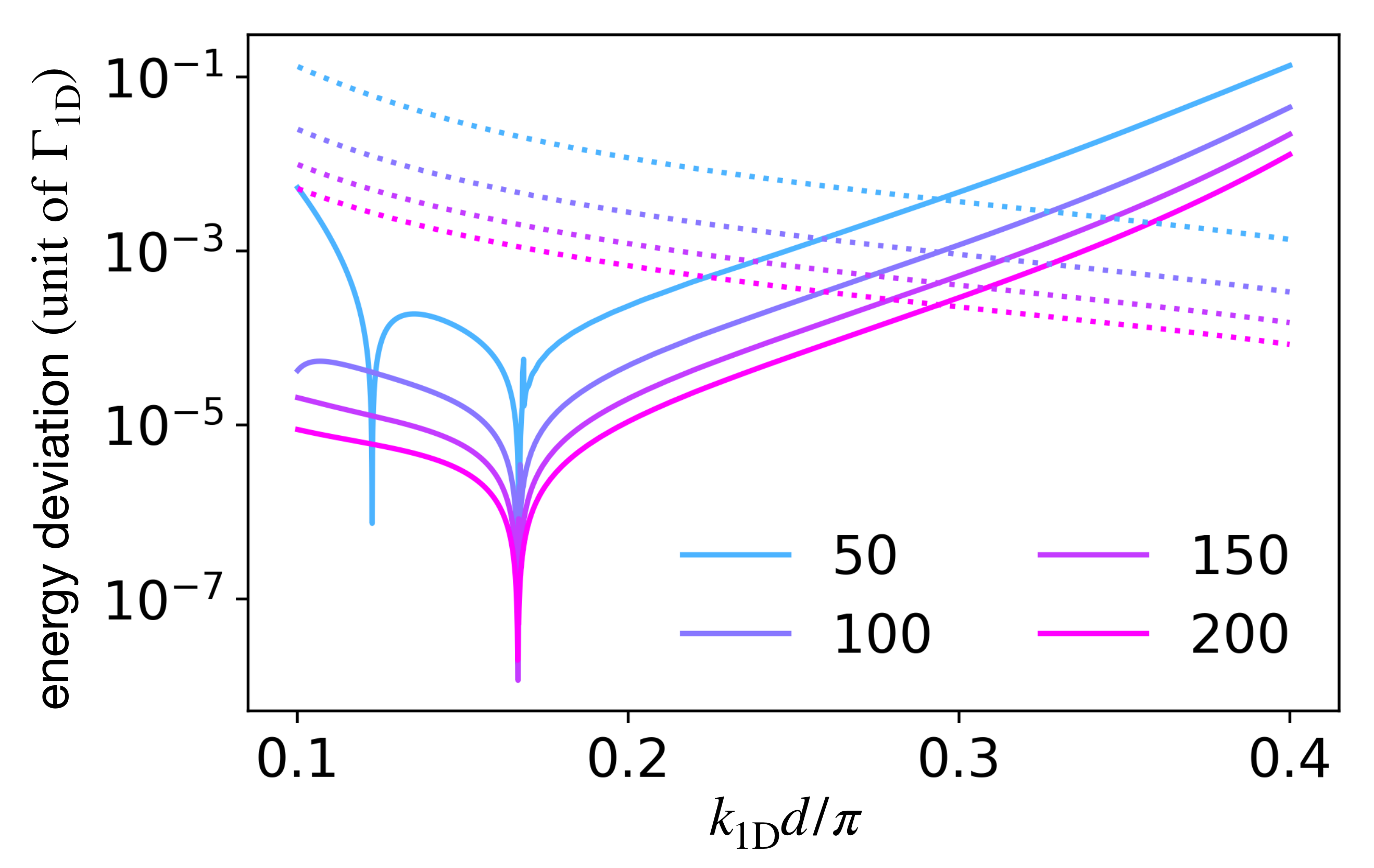}
\caption{The dotted (solid) lines show the deviation between the real part of the eigenvalues (for finite chains) 
and the asymptotic eigenvalues $\omega_{\text{\Rmnum{1}(\Rmnum{2})}}$, as function of $k_{\text{1D}}$
for the type-\Rmnum{1} (type-\Rmnum{2}) dimers for different values of $N$.  }
\label{fig_s2}
\end{figure}

In the limit of infinite chains,
the most subradiant states have vanishing decay rates so that the corresponding eigenvalues are real. Analytical expressions of the asymptotic values can be obtained both from $\mathcal{H}^{K}$ [Eq.~(6) of the main text] and from $\mathcal{H}^{K}_{\text{def}}$ (see Sec.~F).

For the type-\Rmnum{1} dimers,
applying $\mathcal{H}^0$ on $|q\rangle=\sum_{\Delta>0}e^{iq\Delta}|\Delta\rangle$ yields
\begin{equation}
\mathcal{H}^0|q\rangle=\omega_q|q\rangle-\frac{i\Gamma_{\text{1D}}}{2}(g^{0}_{q}|k_{\text{1D}}\rangle+h^{0}_q|{-}k_{\text{1D}}\rangle),
\end{equation}
where the coefficients are
\begin{subequations}
\begin{equation}\label{omega_k}
\omega_q=\frac{\Gamma_{\text{1D}}}{4}\sum_{\epsilon=\pm}\cot(\frac{k_{\text{1D}}+\epsilon q}{2}d),
\end{equation}
\begin{equation}
g^0_{q}=\frac{e^{i(q-k_{\text{1D}})d}}{1-e^{i(q-k_{\text{1D}})d}}+\frac{e^{i(q+k_{\text{1D}})d}-e^{i(q+k_{\text{1D}})Nd}}{1-e^{i(q+k_{\text{1D}})d}},
\end{equation}
\begin{equation}
h^0_q=\frac{e^{i(q+k_{\text{1D}})Nd}}{1-e^{i(q+k_{\text{1D}})d}}.
\end{equation}
\end{subequations}
Suppose that $q$ has a positive imaginary part and $N$ is sufficiently large so that $e^{iqNd}\simeq0$.
This implies that $h^{0}_q=0$ and if we can find a value of $q$ so that $g^0_{q}=0$, the corresponding $|q\rangle$ is an eigenstate of $\mathcal{H}^0$.
For large $N$, this leads to the equation
\begin{equation}
e^{2ik_{\text{1D}}d}=-\frac{1-e^{i(q+k_{\text{1D}})d}}{1-e^{i(q-k_{\text{1D}})d}}.
\end{equation}
The solution is $q=-i\ln\cos(k_{\text{1D}}d)$. Substituting this into the expression
for $\omega_q$ yields the real eigenvalue $\omega_{\Rmnum{1}}=2\cot(k_{\text{1D}}d)$. In Fig.~\ref{fig_s2}, we plot the deviations of energy levels of the dimers from their $N\rightarrow \infty$ asymptotic values.

Counterparts of the type-\Rmnum{2} dimers can be obtained in the same way. One can
also obtain them by using the equivalence between the two types of dimers as presented in Sec.~H
of this Supplemental Material.

\subsection{C. Decay rates of the type-\Rmnum{1} dimers}
\begin{figure}[h]
  \centering
    \includegraphics[width=\textwidth]{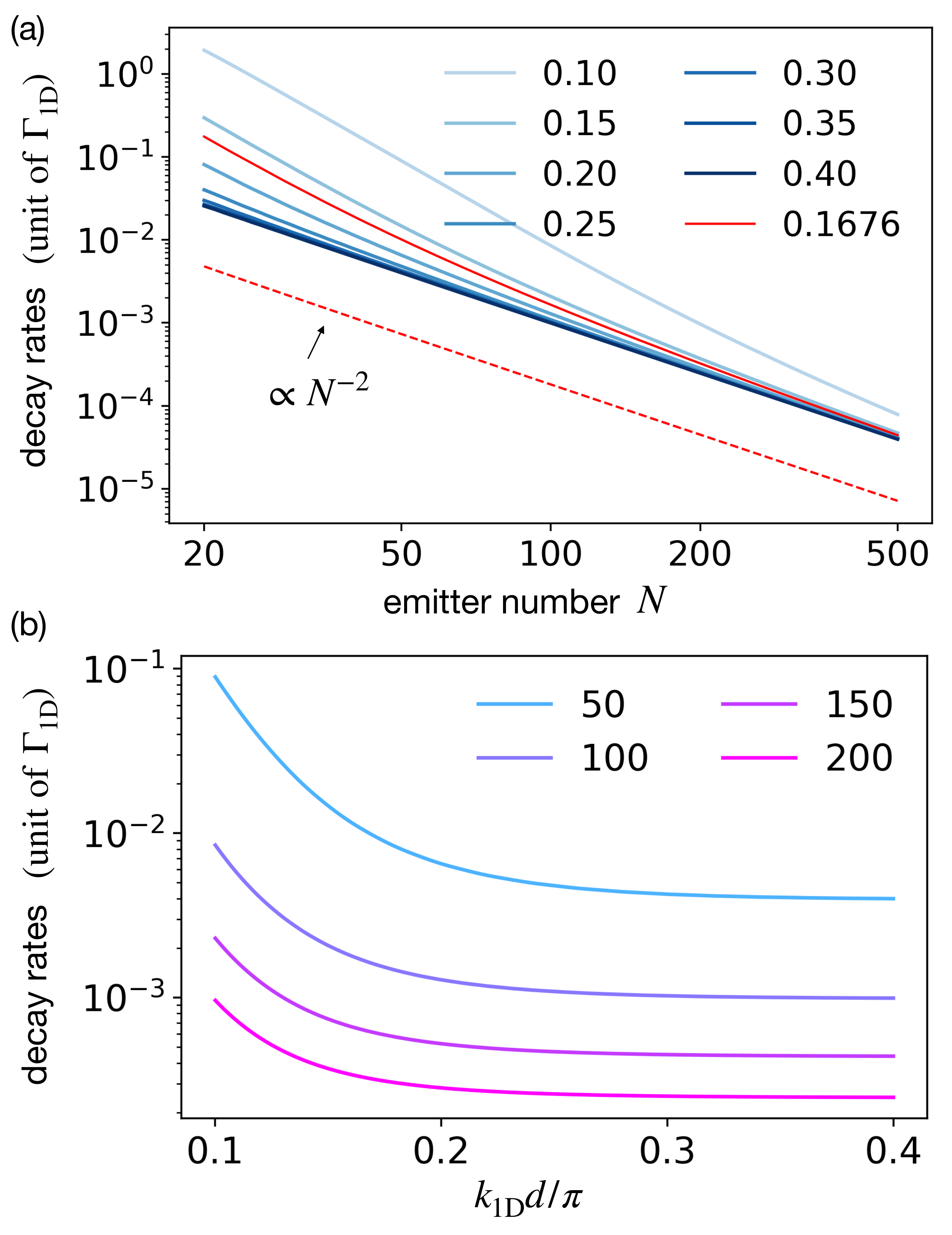}
\caption{Decay rates of the most subradiant type-\Rmnum{1} dimer states.
(a) Decay rates versus emitter number for values of $k_{\text{1D}}d/\pi$ from 0.1 to 0.4 (every 0.05), and 0.1676.
The dashed line is a guide to the eye for the $N^{-2}$ scaling.
(b) Decay rates versus $k_{\text{1D}}$ for fixed $N$.}
\label{fig_s1}
\end{figure}

The decay rates of the most subradiant type-\Rmnum{1} dimer states versus emitter number
$N$ and $k_{\text{1D}}$ are plotted in Fig.~\ref{fig_s1}. Compared with the results for the  type-\Rmnum{2}
states, shown in the main text, the curves are free from dips and wiggles.

\subsection{D. Robustness against spatial disorder}
In this section, we examine the influence of disorder of the spatial positions of the emitters.
It is conceivable that for weak disorder, the eigenstates are only slightly perturbed, while for
more significant disorder, the system may display new physics, such as Anderson localization effects.

\begin{figure}[bp]
  \centering
    \includegraphics[width=\textwidth]{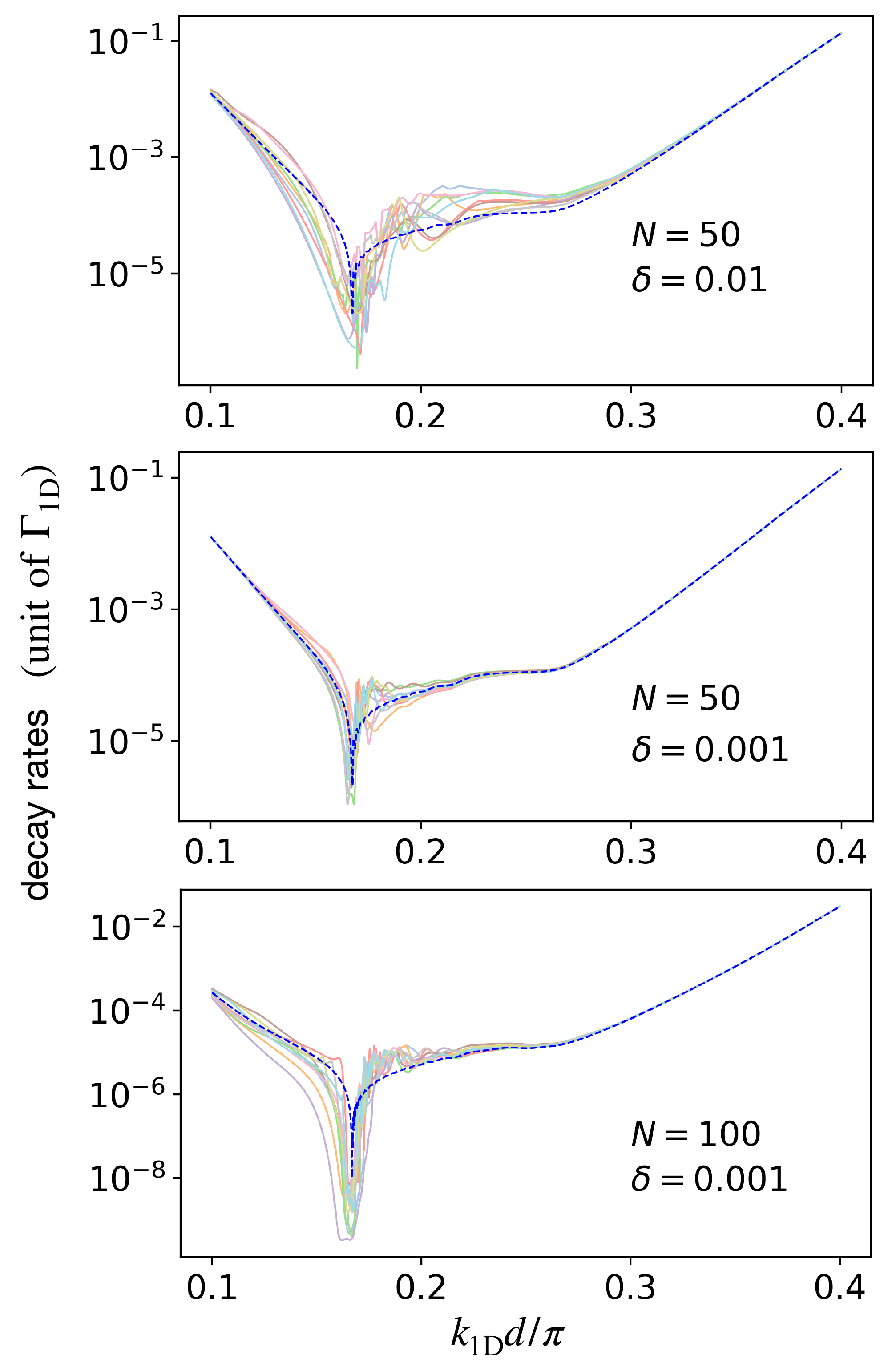}
\caption{
	Effects of spatial disorder on the decay rates of the most subradiant type-\Rmnum{2} dimers, for different values of $N$ and $\delta$ displayed in the figures.
	In each subfigure, the blue dashed line corresponds to the case without disorder, and other lines show results for ten samplings of disordered chains.
}
\label{disorders}
\end{figure}

We restrict ourselves here to the situation of weak disorder.
Then the position of each emitter is assumed to be shifted randomly by a distance uniformly distributed in a small interval
$[-\delta, \delta]d$. We show the decay rates of the
most subradiant type-\Rmnum{2} dimer as a function of $k_{\text{1D}}d$ in Fig.~\ref{disorders} for a number of random realizations of the disorder.
Our simulations show that the dip near $k_{\text{1D}}d=\pi/6$ is robust against the disorder, and that disorder can even lead to further suppression of the decay rates.

\subsection{E. Full expression of terms omitted in Eq.~(5) of the main text}
The omitted ``tails'' are expressed as
\begin{equation}
\begin{aligned}
\text{(tails)}=& \frac{i\Gamma_{\text{1D}}}{2}e^{i(k_{\text{1D}}+\frac{K}{2})\Delta}\sigma^{\dagger}_{-k_{\text{1D}}}(\sigma^{\dagger}_{K+k_{\text{1D}}})_{R}|G\rangle \\
& +\frac{i\Gamma_{\text{1D}}}{2}e^{i(k_{\text{1D}}-\frac{K}{2})\Delta}\sigma_{k_{\text{1D}}}^{\dagger}(\sigma_{K-k_{\text{1D}}}^{\dagger})_L |G\rangle,
\end{aligned}
\end{equation}
where $|G\rangle$ is the state where no emitters are excited,
$\sigma_{p}^{\dagger}=\sum_{m}e^{ipz_m}\sigma_m^{\dagger}$, and the foot indices $R$ and $L$ restrict the summation over sites to the intervals $(z_N-\Delta, z_N]$ and $[z_1, z_1+\Delta)$, respectively. For dimer states dominated by small values of $\Delta$, the ``tail'' terms
are well restricted to the ends of the chain.

\subsection{F. Mapping from $\mathcal{H}^{K}$ to $\mathcal{H}^{K}_{\text{def}}$}

The one-to-one correspondence between the eigenstates of $\mathcal{H}^{K}$ and the even eigenstates of $\mathcal{H}^{K}_{\text{def}}$
can be understood from the following simple observation.

From Eq.~(6) of the main text, we see that the summation over $\epsilon'=\pm1$ in $\mathcal{H}^{K}$ can be formally represented by writing
$(\mathcal{H}^{K})_{\Delta,\Delta'}=A^{+}_{\Delta,\Delta'}+A^{-}_{\Delta,\Delta'}$, where
\begin{equation}
A^{\pm}_{\Delta,\Delta'}=-\frac{i}{2}\Gamma_{\text{1D}}\sum_{\epsilon=\pm1}e^{i(k_{\text{1D}}+\epsilon\frac{K}{2})|\Delta-(\pm)\Delta'|}.
\end{equation}
While these quantities are introduced for application to the case of $\Delta, \Delta' > 0$,
they formally obey the symmetry, $A^{+}_{\Delta,\Delta'}=A^{-}_{\Delta,-\Delta'}$ and, hence
the action of $\mathcal{H}^{K}$ on a general state obeys the following set of equations,
\begin{equation}
\begin{aligned}
&\sum_{\Delta'>0}(\mathcal{H}^{K})_{\Delta,\Delta'}\langle\Delta'|\psi\rangle \\
= &\sum_{\Delta'>0} A^{+}_{\Delta,\Delta'}\langle\Delta'|\psi\rangle+\sum_{\Delta'>0} A^{-}_{\Delta,\Delta'}\langle\Delta'|\psi\rangle \\
= &\sum_{\Delta'>0} A^{+}_{\Delta,\Delta'}\langle\Delta'|\psi_{\text{def}}\rangle+\sum_{\Delta'<0} A^{+}_{\Delta,\Delta'}\langle-\Delta'|\psi_{\text{def}}\rangle.
\end{aligned}
\end{equation}

In the last line we introduce the even states, $|\psi_{\text{def}}\rangle$, defined for both positive
and negative $\Delta$, and satisfying $\langle\Delta|\psi\rangle=\langle\Delta|\psi_{\text{def}}\rangle$
for $\Delta>0$, and we observe that the last expressions can be combined in a single
sum  $\sum_{\Delta' \neq 0}  2(\mathcal{H}^{K}_{\text{def}})_{\Delta,\Delta'} \langle\Delta'|\psi_{\text{def}}\rangle$, where
\begin{equation}
(\mathcal{H}^{K}_{\text{def}})_{\Delta,\Delta'}=\frac{1}{2} A^{+}_{\Delta,\Delta'}.
\end{equation}

\subsection{G. Defect-induced localized subradiant states}
We denote the effective Hamiltonian of a chain with the $m^{th}$ emitter missing by $H_{-m,\text{def}}$.
This defect separates the chain into a left and a right subchain, where Bloch one-excitation states, $|q_{L}\rangle$ and $|q_{R}\rangle$,
are defined as
\begin{equation}
|q_{L(R)}\rangle=\sum_{m\in L(R)}e^{iqz_m}|m\rangle.
\end{equation}
Then we have
\begin{subequations}\label{sp h-m}
\begin{equation}
\begin{aligned}
H_{-m,\text{def}}|q_{L}\rangle= &{\ } \omega_q|q_{L} \rangle-\frac{i\Gamma_{\text{1D}}}{2}\bigg(g_{L,q}|k_{\text{1D};L}\rangle \\
+\beta_{q} & |k_{\text{1D};R}\rangle-h_{L,q}|{-}k_{\text{1D};L}\rangle\bigg),
\end{aligned}
\end{equation}
\begin{equation}
\begin{aligned}
H_{-m,\text{def}}|q_{R}\rangle= &{\ } \omega_q|q_{R}\rangle+\frac{i\Gamma_{\text{1D}}}{2}\bigg(h_{R,q}|{-}k_{\text{1D};R}\rangle\\
-\theta_{q} & |-k_{\text{1D};L}\rangle -g_{R,q}|k_{\text{1D};R}\rangle\bigg),
\end{aligned}
\end{equation}
\end{subequations}
where the coefficients are given as
\begin{subequations}
\begin{equation}
g_{L,q}=\frac{e^{i(q-k_{\text{1D}})z_1}}{1-e^{i(q-k_{\text{1D}})d}},\quad
g_{R,q}=\frac{e^{i(q-k_{\text{1D}})(z_m+d)}}{1-e^{i(q-k_{\text{1D}})d}},
\end{equation}
\begin{equation}
h_{L,q}=\frac{e^{i(q+k_{\text{1D}})z_m}}{1-e^{i(q+k_{\text{1D}})d}},\quad
h_{R,q}=\frac{e^{i(q+k_{\text{1D}})(z_N+d)}}{1-e^{i(q+k_{\text{1D}})d}},
\end{equation}
\begin{equation}
\beta_q=\frac{e^{i(q-k_{\text{1D}})z_1}-e^{i(q-k_{\text{1D}})z_m}}{1-e^{i(q-k_{\text{1D}})d}},
\end{equation}
\begin{equation}
\theta_q=\frac{e^{i(q+k_{\text{1D}})(z_m+d)}-e^{i(q+k_{\text{1D}})(z_N+d)}}{1-e^{i(q+k_{\text{1D}})d}}.
\end{equation}
\end{subequations}
The expression of $\omega_q$ is identical to that of Eq.~\eqref{omega_k}. We expect that the eigenvalues
will be expressed by $\omega_q$ for some specific values of $q$ with contributions to the eigenstates from the degenerate states $|\pm q_{L}\rangle$
and $|\pm q_{R}\rangle$. Indeed, one verifies by inspection that a 
superposition, $c_L|\psi_q\rangle_L+c_R|\psi_q\rangle_R$, of $|\psi_q\rangle_L\propto g_{L,-q}|q_{L}\rangle
-g_{L,q}|-q_{L}\rangle$ and $|\psi_q\rangle_R\propto h_{R,-q}|q_{R}\rangle- h_{R,q}|-q_{R}\rangle$,
lead to cancellation of the
$|\pm k_{\text{1D};L}\rangle$ and $|\pm k_{\text{1D};R}\rangle$ terms in Eq.~\eqref{sp h-m}
and that the coefficients $c_L$ and $c_R$ can be found if the determinant of the following matrix vanishes:
\begin{equation}
\begin{pmatrix}
& g_{L,q}h_{L,-q}-g_{L,-q}h_{L,q} & \theta_q h_{R,-q}-\theta_{-q}h_{R,q} \\
& \beta_{q}g_{L,-q}-\beta_{-q}g_{L,q} & g_{R,q}h_{R,-q}-g_{R,-q}h_{R,q}
\end{pmatrix}
\end{equation}
This condition is further evaluated to be
\begin{equation}\label{sss}
\begin{aligned}
\sum_{\epsilon=\pm}\frac{1}{A_{\epsilon q}^2} & e^{-i\epsilon qd(N-1)}-\frac{e^{2ik_{\text{1D}}d}}{A_q A_{-q}}
\sum_{\epsilon=\pm}e^{i\epsilon qd(N-1)} \\
& =\frac{1-e^{2ik_{\text{1D}}d}}{A_q A_{-q}}
\sum_{\epsilon=\pm}e^{i\epsilon qd(N-2m+1)},
\end{aligned}
\end{equation}
where
$$A_{q}=2-e^{i(q-k_{\text{1D}})d}-e^{-i(q-k_{\text{1D}})d}.$$
It is expected that the solution for $q$ has positive
imaginary part, hence $e^{iqNd}\approx 0$. Then Eq.~(\ref{sss}) can be evaluated to
\begin{equation}\label{sp_solution}
\begin{aligned}
\frac{A_{-q}}{A_{q}}- & e^{2ik_{\text{1D}}d}= (1- e^{2ik_{\text{1D}}d})\\
& \times [(\cos k_{\text{1D}}d)^{2(N-m)}+(\cos k_{\text{1D}}d)^{2(m-1)}].
\end{aligned}
\end{equation}
When the missing site is far from the chain ends so that
the right hand side of the above equation can be ignored, we have
\begin{equation}
e^{2ik_{\text{1D}}d}=A_{-q}/A_q,
\end{equation}
with the solution $q_{\text{\Rmnum{1}}}=-i\ln\cos(k_{\text{1D}}d)$. Similarly for the case corresponding to
the type-\Rmnum{2} states, we obtain $q_{\text{\Rmnum{2}}}=\pi-0.5i\ln\cos(2k_{\text{1D}}d)$.
Substituting these expressions into $\omega_q$ yields the asymptotic eigenvalues.

Terms on the right hand side of Eq.~(\ref{sp_solution}) are suppressed exponentially in
the length $\min N_{LR}$ of the shorter subchain (left or right side of the missing site). The coupling at the end provides a
correction, $q=q_{\text{\Rmnum{1}}}+\delta$, with
\begin{equation}
\delta=\frac{-i}{2d}\cot^2(k_{\text{1D}}d)(\cos k_{\text{1D}}d)^{\min N_{LR}}.
\end{equation}
Substituting this into the expression for $\omega_q$ yield an exponentially suppressed decay rate
of the localized state as function of $\min N_{LR}$.

The localized states are exponentially suppressed at the chain ends and the eigenstates are approximately given in the form of
$|q_{\text{\Rmnum{1}};R}\rangle+|-q_{\text{\Rmnum{1}};L}\rangle $
as we present in the main text.

\subsection{H. Properties of $\mathcal{H}^{K=\pi/d}_{\text{def}}$ for the type-\Rmnum{2} dimers}

We have,
\begin{equation}
(\mathcal{H}^{\pi/d}_{\text{def}})_{\Delta,\Delta'}=-i\frac{1}{4}\Gamma_{\text{1D}}\sum_{\epsilon=\pm1}e^{i(k_{\text{1D}}d+\epsilon\frac{\pi}{2})|\Delta-\Delta'|},
\end{equation}
where the indices $\Delta$ and $\Delta'$ have been transformed to dimensionless integers for convenience of notation.

If $\Delta$ and $\Delta'$ have opposite parity, i.e., one is even and the other is odd, $|\Delta-\Delta'|$ will be odd
and consequently
\begin{equation}
e^{i\frac{\pi}{2}|\Delta-\Delta'|}+e^{-i\frac{\pi}{2}|\Delta-\Delta'|}=0.
\end{equation}
It means that $(\mathcal{H}^{\pi/d}_{\text{def}})_{\Delta,\Delta'}=0$, i.e., the odd and
even $\Delta$ are not coupled. Thus we can write
\begin{equation}
\mathcal{H}^{\pi/d}_{\text{def}}=\mathcal{H}^{\pi/d}_{\text{def};\,odd}+\mathcal{H}^{\pi/d}_{\text{def};\,even},
\end{equation}
where the odd and even terms commute.

The subchain consisting of all odd sites
$\Delta=\pm1, \pm3,\cdots$ does not couple to the defect at $\Delta=0$ and provides no localized solutions.
It is therefore sufficient to consider $\mathcal{H}^{\pi/d}_{\text{def};\,even}$, acting on the even sites $\Delta=\pm2, \pm4,\cdots$. By substituting
$\Delta=2\xi$ ($\xi=\pm1, \pm2, \cdots$) into $\mathcal{H}^{\pi/d}_{\text{def};\,even}$, we find
\begin{equation}
\begin{aligned}
(\mathcal{H}^{\pi/d}_{\text{def};\,even})_{\xi,\xi'}= & -i\frac{1}{4}\Gamma_{\text{1D}}\sum_{\epsilon=\pm1}
e^{i(2k_{\text{1D}}d+\epsilon\pi)|\xi-\xi'|} \\
=& -i\frac{1}{2}\Gamma_{\text{1D}}e^{i2k_{\text{1D}}d|\xi-\xi'|}(-1)^{|\xi-\xi'|}\\
=& -i\frac{1}{2}\Gamma_{\text{1D}}e^{i2k_{\text{1D}}d|\xi-\xi'|}(-1)^{\xi+\xi'}.
\end{aligned}
\end{equation}
Using a local phase transformation, $|\xi\rangle\rightarrow (-1)^{\xi}|\xi\rangle$, the above expression can be transformed to
\begin{equation}
(\mathcal{H}^{\pi/d}_{\text{def};\,even})_{\xi,\xi'}\rightarrow
-i\frac{1}{2}\Gamma_{\text{1D}}e^{i2k_{\text{1D}}d|\xi-\xi'|},
\end{equation}
which is equivalent to the Hamiltonian for the 
type-\Rmnum{1} $(K=0)$ dimers with the scaled parameter $2k_{\text{1D}}d$.

To summarize, the localization of $\mathcal{H}^{\pi/d}_{\text{def}}(k_{\text{1D}}d)$, which corresponds to the
type-\Rmnum{2} dimers, is
equivalent to that of $\mathcal{H}^{0}_{\text{def}}(2k_{\text{1D}}d)$ which belongs to the
type-\Rmnum{1} dimers.

\subsection{I. Phase profiles of the localized states}

\begin{figure}[b]
  \centering
    \includegraphics[width=\textwidth]{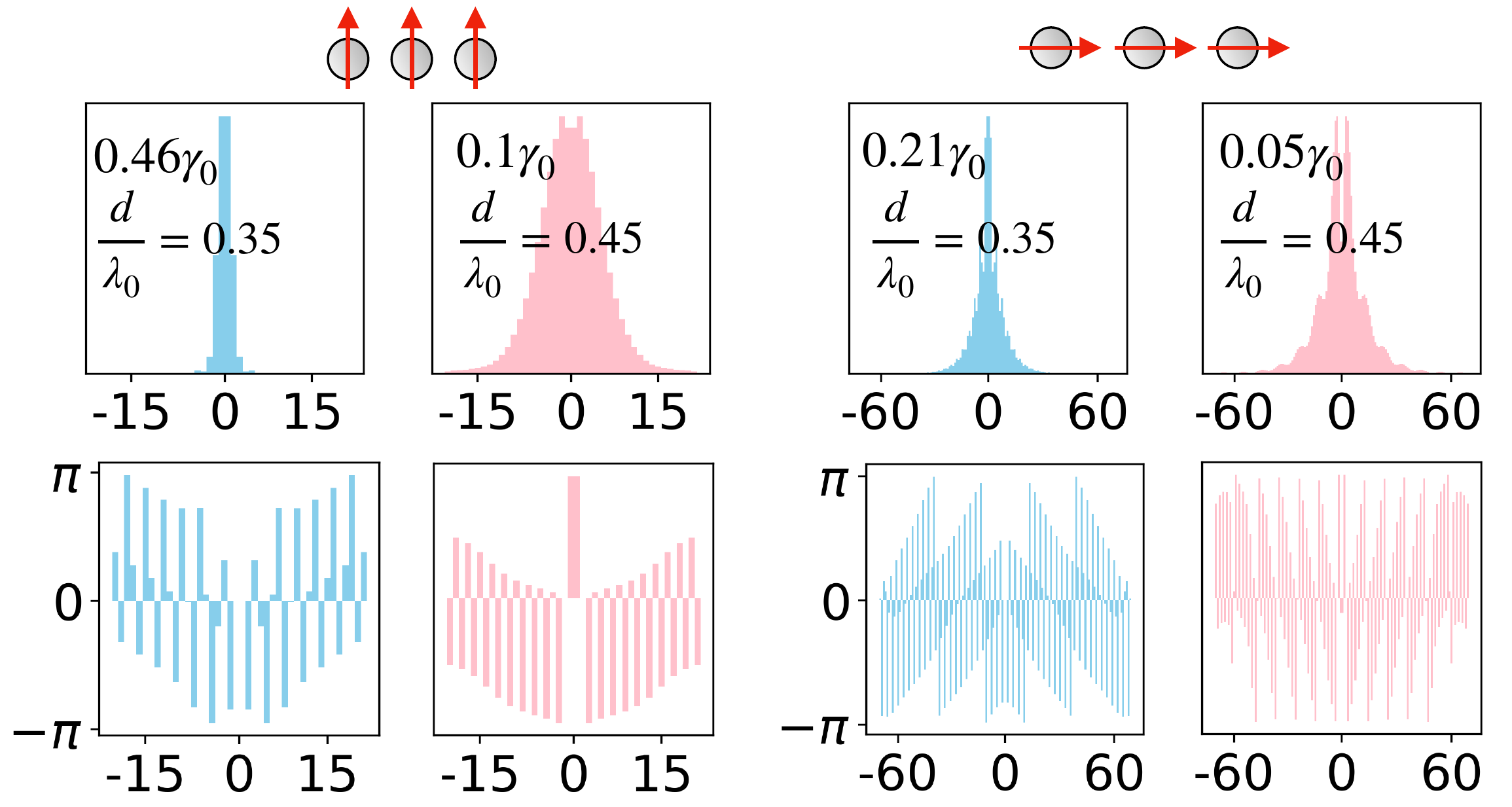}
\caption{ Bottom panels: phase profiles of the localized states discussed in the main text.}
\label{sp_phases}
\end{figure}

For the 1D waveguide case, the phase of the localized state is uniform for $0 < k_{\text{1D}}d < \pi/2$, and
flips by $\pi$ per site if $\pi/2 < k_{\text{1D}}d < \pi$. For coupling to the 3D free space vacuum field,
the amplitude and phase profiles are not as regular as in the 1D waveguide case. In Fig.~\ref{sp_phases}
we show the phase profiles of the localized states illustrated in Fig. 4(d,e) of the main text (the
amplitude profiles are replicated for convenience). 

In the figure, the emitter number is not
specified, because profiles of the localized states are almost the same, as long as their widths are 
adequately shorter than the chain lengths.

\end{document}